# Volumetric 3D-printed antennas, manufactured via selective polymer metallization

Dmitry Filonov, Sergey Kolen, Andrey Shmidt,
Yosi Shacham-Diamand, Amir Boag, and Pavel Ginzburg

*Abstract* - **Additive manufacturing paves new ways to efficient exploration of the third space dimension, providing advantages over conventional planar architectures. In particular, volumetric electromagnetic antennas can demonstrate superior characteristics, outperforming their planar counterparts. Here a new approach to fabrication of electromagnetic devices is developed and applied to antennas, implemented on curved surfaces. Highly directive and broadband antennas were 3D-printed on hemispherical supports. The antenna skeleton and the support were simultaneously printed with different polymer materials - PLA mixed with graphene flakes and pure PLA, respectfully. Weakly DC-conductive graphene PLA-based skeleton was post-processed and high quality conductive copper layer was selectively electrochemically deposited on it. The antenna devices were found to demonstrate radiation performance, similar to that achievable with conventional fabrication approaches. However, additive manufacturing of RF antennas provides superior capabilities of constructing tailor-made devices with properties, pre-defined by non-standardized end users.**

## I. Introduction

A significant effort has been devoted to developing various antenna devices, since these components are an essential part of any wireless communication system in a broad sense [1],[2]. A variety of different architectures has been demonstrated over the years and came to address specific requirements, demanded by system specifications. Main antenna parameters are achieved by finding and optimizing a specific geometry, which should have a sufficient number of degrees of freedom in order to ensure a desired solution. While three-dimensional (3D) architectures can provide significantly larger search sets, 2D implementations are usually more favourable choice owing to their significantly higher integrability with printed electronic circuitry. However, newly emerging Additive Manufacturing technologies suggest reconsidering the volumetric approaches and explore the third space dimension in integrated antenna applications. Additive Manufacturing is rapidly advancing, as the demand for low cost, compact, complex shape, light weight, time efficient in production, and environmentally friendly technologies continues to expand over many different technological areas [3], including radio frequency (RF) applications.

Nowadays, there are many different Additive Manufacturing techniques, which allow production of high quality RF devices [4]. Existent approaches include and not limited to 3D manufacturing with CNC milling technique [5],[6], Laser Direct Structuring (LDS) [7],[8],[9], fabrication by conformal printing of metallic inks [10], 3D printing by conductive inkjet printing [11], ultrasonic wire mesh embedding [12], and metal deposition trough a mask on a curve surface [13], [14]. While the vast majority of the beforehand mentioned techniques require involvement of quite an expensive and maintenance demanding machinery, recent advances and subsequent cost reductions of extrusion-based polymer printers suggest considering them as an affordable tool for manufacturing of tailor-made antenna devices. Many different polymer materials have been explored so far and include PLA, ABS, PETG, different alloys and many others. For example, printed plastic materials were shown to be integrated within antenna design (e.g. [15],[16]) and additive manufacturing of low-profile devices with multi-materials has been demonstrated [17]. In many cases, however, device supports have to be manufactured separately from their functional (conductive) parts. As a result, accurate printing process requires time-consuming alignments to guarantee high quality results. This process is even more difficult when conductive parts need to be printed on geometrically complicated 3D substrates, often described by the B-splines.

Here we demonstrate a new technique, which relies on simultaneous printing of antenna skeletons with DC conductive polymer and supports, made of low dielectric contrast isolating plastic. The structure includes conducting interconnect which is required for the copper plating step. Unlike traditional machining techniques, which create an object by drilling or etching to remove and shape materials, our three-dimensional objects are generated by a computer model, which is subsequently implemented via layering of different functional materials. This not only addresses the alignment difficulties for the conventional printing applications, but also provides numerous opportunities for the fabrication of a wide range of different electromagnetic structures with integrated conducting elements. Although the proposed technique requires a subsequent (e.g. post-printing) electrochemical post-processing, this extra step is rather simple, requiring only a temperature controlled bath, current source and can be batch processed for high throughput manufacturing. This plating step should be taken into account at the design stage, as will be elaborated hereafter. Disadvantages of the extrusion-based method are moderate surface roughness and probable voids inside 3D printed structures. While those parameters might have an impact on high-frequency applications, low GHz realizations are tolerant to mm-scale imperfections. Furthermore, the main required feature from the graphene-based skeleton is the low frequency conductivity. This parameter is achieved by a mesh of electrically connected filaments. After the first electrochemically deposited layer, the internal structure of the skeleton does not play a significant role, as a continuous smooth metal film forms. This Letter is organized as follows: 3D printing and selective metallization approach is presented first and then followed by characterization of two antennas, fabricated on hemispherical supports. Directive resonant Yagi-Uda and broadband Archimedean spiral antennas demonstrate the capabilities of the new manufacturing methodology.

## II. Additive manufacturing of RF antennas with selective skeleton metallization

Antenna manufacturing procedure consists of two main steps – 3D printing of the structures and selective electrochemical deposition of conducting layers. 3D printer (BCN3D Sigma R17) has two independent dual extruders, which allow simultaneous printing with two different materials. The machine automatically interchanges the extruders, when different material sections are involved. The 0.3 mm diameter nozzles were chosen in order to strike the balance between the quality of printed surfaces and the running time. Among a variety of different possible materials, non-conducting PLA (it also has low high frequency losses, as it will be shown hereafter) and its conducting version, mixed with graphene flakes were used. It is worth noting that graphene PLA (G-PLA, commercial 3D Lab Graphene Conductive PLA) has moderate volume resistivity of 3.8 Ω·cm, which by itself is insufficient to support high frequency currents, essential for efficient radiation of electromagnetic waves (DC conductivity was measured with a multimeter, assuming a uniform current in cross-section. The measured values agree with the vendor data). However, DC conductivity of this material allows performing galvanic electrochemical deposition (e.g. electroplating) of high quality copper layers with bulk resistivity in the range of 1.8-2 μΩ·cm. G-PLA is conducting (i.e. enough to allow plating), while the PLA is an insulator, thus the plating occurs only on the G-PLA and only on parts, which are connected to the plating power supply allowing a closed circuit. Therefore, it is a design and printing issue that should be taken into consideration. Note that the G-PLA conductivity should be low enough and/or the plating current should be limited so the serial voltage drop on the G-PLA during the plating process. The DC-resistivity of the metalized G-PLA was found to be three-four orders of magnitude higher (~24 mΩ·cm) than its pure version, while the PLA support remained nonconductive.

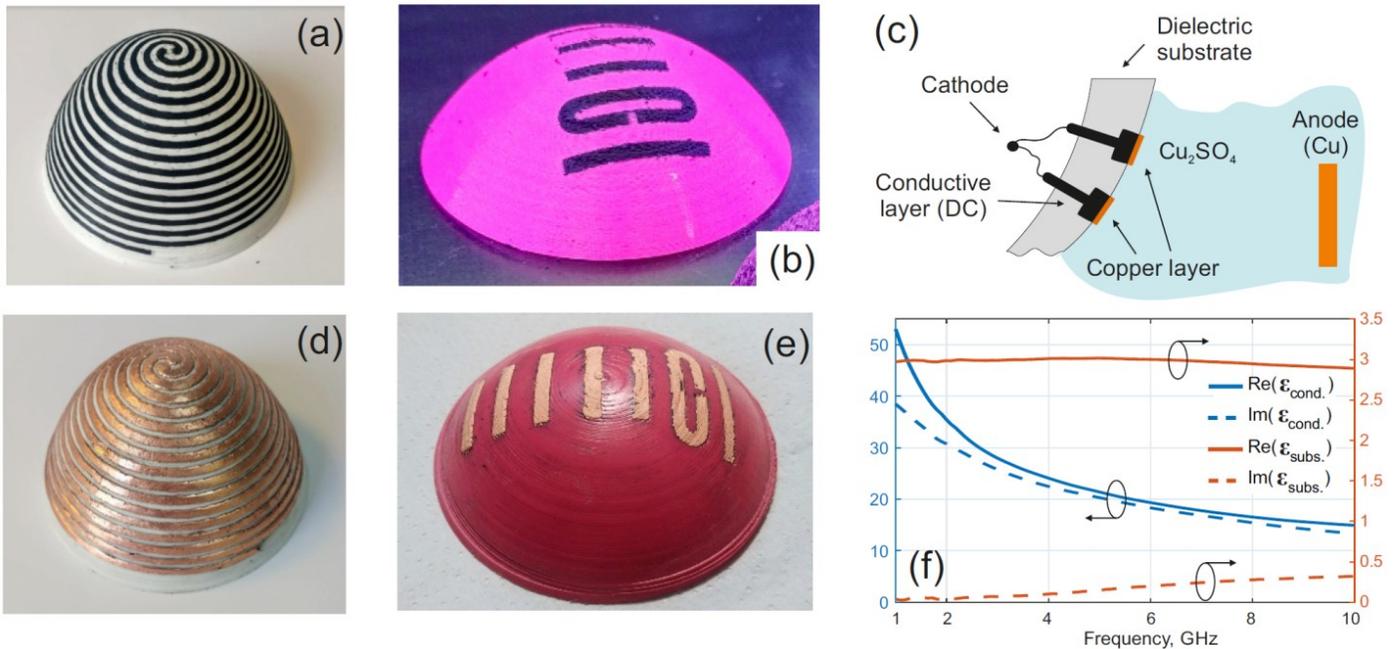

Fig. 1. Additive manufacturing of volumetric antenna devices. Photographs of 3D printed antennas: (a), (d) broadband Archimedean spiral; (b), (e) directive Yagi-Uda antenna - before and after the metallization process (copper layer is deposited on the antenna skeleton), respectively. Black lines are the antenna skeleton, made of PLA mixed with graphene flakes. Antenna supports are made of nonconductive PLA (white color – panel (a,d), magenta – panel (b,e). The color has no impact on the electromagnetic properties). (c) Schematics of a cut, demonstrating the layers structure of the devices. (f) Material parameters of PLA materials. Nonconductive PLA – orange curves, right vertical axis. Graphene PLA – blue curves, left vertical scale. Solid lines – real, dashed lines imaginary part of the relative permittivities.

The fabrication process appears in Fig. 1. Panels (a) and (b) demonstrate the first printing step. Hemispherical supports were implemented with either white or magenta PLAs (coloration of materials is provided by the vendor and does not influence electromagnetic properties of the structure). Hemispherical geometry was chosen for demonstrating the capabilities of 3D manufacturing and is not related to a specific application. G-PLA has a black colour and antennas skeletons can be clearly seen on the photograph. The skeleton is embedded into the support (Fig. 1(c)), demonstrating the capability of the simultaneous printing with several materials. At the end of the printing (the first manufacturing stage), the samples were post-processed with acetone in order to improve the quality of the surfaces. Low GHz applications (~10 GHz and less) can tolerate mm-scale roughness without significant degradation of electromagnetic performance.

The second stage of the antennas' fabrication includes the galvanic electrochemical deposition at room temperature (~22-24 °C). The samples are placed into a saturated (estimated by observing the undissolved residue) Copper sulfate ($CuSO_4$) solution and 4% sulfuric acid (Sigma Aldrich). Copper wires are added into the jar in order to replenish the metal ions' concentration in the electrolyte. G-PLA skeleton serves as a cathode, while the coper wire is the anode. It is worth noting that G-PLA surfaces should be purified with ethanol prior to the deposition process. The used DC current was 1 mA/cm². The skeleton can be divided into several independent sections or all the antenna elements can be connected with auxiliary conducting bridges, which are removed at the end of the process (Fig. 1(c)). The overall deposition time was ~24 hours. The thickness of the metalized layer was found to be ~0.7 mm, which exceeds the skin depth of copper at the frequencies of interest by a factor of ~1000 (the skin depth of copper at the considered frequency range is on the micron scale). Voids inside G_PLA material were found to have a minor impact on the metallization process. The final antenna layouts appear in Fig. 1 (d and e).

Electromagnetic parameters of the PLA materials were retrieved in order to perform the design and optimization. It is worth mentioning that permittivities of those nonmagnetic 3D-printed composites depend on the volumetric filing factor, which is predefined and can be further controlled with the printer software. Here, the factors to be considered are mechanical rigidity, overall weight, printing time and (in the case of G-PLA) the DC conductance. The dielectric constant and loss tangent of the printed materials were measured in 1-10 GHz frequency range by using Keysight high-temp dielectric probe (85070D). The resulting dispersion curves of the materials' permittivity appear in Fig. 1(f). Samples were polished before performing the tests. Furthermore, the retrieval procedure was repeated with waveguide geometry (WG16, X band) and an agreement within 10% were found, verifying the validity of the results. Pure PLA has a relatively low loss permittivity ($\varepsilon_{sub}$) with a weak dispersion, which simplifies the design of broadband antenna elements (loss tangent is 0.0012 - 0.11 over the whole measured frequency band). On the other hand, graphene PLA has strong material dispersion and significant material losses – the loss tangent approaches unity (0.72-0.89).

This behavior corresponds well to the fact that this material has modest DC conductivity. However, the relative fraction of the conductive skeleton in the overall structure is rather small and the electromagnetic field overlap with the lossy dispersive material is not high (copper layer screens the G-PLA skeleton), which is essential for maintaining antenna performances.

### III. CHARACTERIZATION OF 3D PRINTED ANTENNAS

Two types of antennas were manufactured in order to demonstrate the capabilities of the developed method. The first one is the Archimedean spiral broadband antenna. It consists of a long (150 cm) continuous strip, wrapped around the hemisphere. This geometry is typically employed in RFID and other applications and was chosen here to demonstrate the metallization capabilities of large connected surface areas with nontrivial geometry. The second demonstration is the resonant Yagi-Uda antenna, which is employed in a broad range of applications, where device directivity is the main parameter. Here, this geometry was exploited in order to demonstrate that resonant properties of elements are not degraded owing to the additive manufacturing process. Antenna designs were performed with CST Microwave Suit, by taking into account the structures layouts and retrieved material parameters of the constitutive components (Fig. 1(f)).

*Archimedean Spiral Antenna*

The final layout of the broadband antenna appears in Fig. 1(d) (Technical Information section includes the detailed geometrical parameters). The antenna was implemented on PLA hemispherical shell with a radius of 50 mm and 6 mm thickness. Two identical spirals, rotated one with respect to the other by 180 degrees, contain 8 full turns. The line spacing was chosen to be 2 mm and the metalized line width is also 2 mm (e.g. [18] for the design guidelines). Those parameters were obtained after performing a set of numerical optimizations towards achieving the broadband operation of the device. Performances of the antenna are summarized in Fig. 2. Panel (a) shows the broadband matching on the level of –(7-10) dB, which is quite typical for spiral antennas. Insets to panel (a) show the radiation patterns (linear scale) at the corresponding frequencies. Cuts of those rotationally symmetric patterns are given in panels (b, c, d). The shapes maintain their parameters (directivity, beam width, side lobs) over the broad frequency range in the full agreement with the main specification of this antenna. Numerical calculation agrees well with the experimental data (patterns are normalized to the global maximum). It can be seen that the differences between the numerical and experimental results become more pronounced at higher frequencies (Fig.

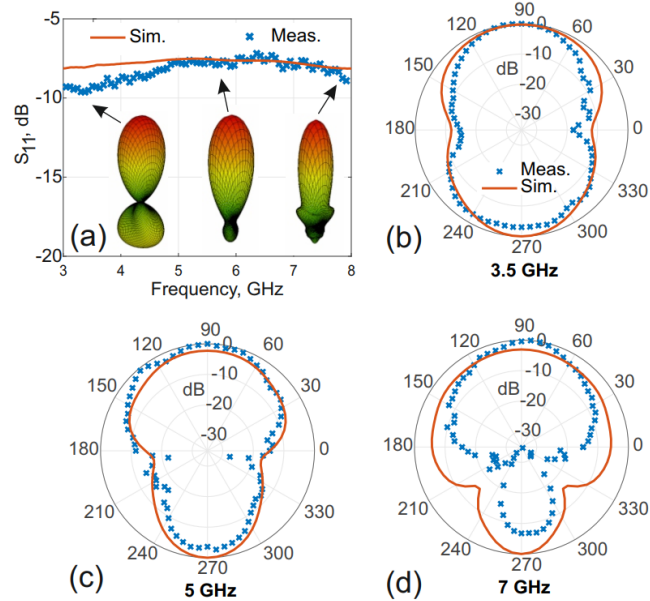

2(d)), since the surface roughness starts playing a role.

Fig. 2. Performance of the 3D printed spiral broadband antenna. (a) Absolute value of $S_{11}$, as the function of frequency. Inset – numerically calculated radiation patterns (linear scale) at different frequencies, indicated with arrows. Cuts of the normalized far-field patterns (logarithmic scale) at: (b) 3.5GHz, (c) 5 GHz, (d) 7 GHz. Red solid lines – numerical simulation. Blue lines/dots – experimental data.

*Yagi-Uda Antenna*

The final layout of the Yagi-Uda antenna appears in Fig. 1(e). Geometrical details are given in the Section V. The width of all of the elements is 3.5 mm, and thickness is 3 mm. The antenna was implemented on PLA spherical shell with radius of 44 mm, height - 24 mm and 6 mm thickness.

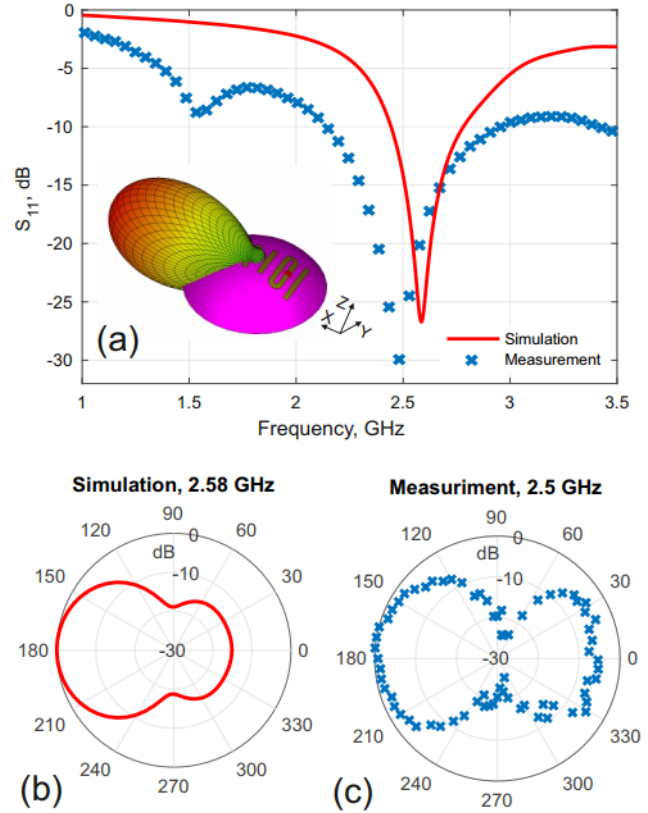

Fig. 3. Performances of 3D printed Yagi-Uda antenna. (a) Absolute value of $S_{11}$, as a function of frequency. Inset – numerically obtained directivity pattern, linear scale. The cut of the normalized far-field pattern at: (b) 2.58 GHz, numerical data. (c) 2.5 GHz, experimental data. Red solid lines – numerical simulation. Blue lines/dots – experimental data.

The measured $|S_{11}|$ profile of the Yagi-Uda antenna is compared with the numerical analysis in Fig. 3(a). The resonant frequency of the structure appears at 2.5 GHz, which deviates by 90 MHz from targeted working frequency (equivalent to 3.6% error). This type of shift is relatively common in the case of resonant elements and originate from uncertainties in permittivities and exact geometrical parameters of material components. Additional iterative engineering effort is done in the case of commercial products, where nominal numbers play significant roles. Note, that small frequency shift can be also observed in the case of the broadband antenna (Fig. 2(a)), nevertheless it is much more difficult to quantify them owing to the low quality factor of the structure.

The Yagi-Uda directivity pattern (linear scale) appears in the inset to Fig. 3(a) and its cut (logarithmic scale) is presented in panels (b) and (c) for numerical and experimental cases, respectively. Clear resonant behavior can be observed. Theoretical gain of a flat Yagi-Uda structure with five directors is 11 dBi [19],[20], while in our 3D model the highest optimized gain is 8.28 dBi with the main radiation direction along x-axis (see the inset). Experimental values of the gain (not measured) are expected to overestimate this value by 10-20%. The main reasons for this degradation is the curved structure of Yagi-Uda antenna support and lossy material components, involved in the manufacturing process.

## IV. OUTLOOK AND CONCLUSIONS

Additive manufacturing methodology for the production of volumetric RF antennas was developed and demonstrated on two different devices. The concept is based on simultaneous 3D printing of two different materials – antenna skeleton, made of a weakly DC conducting material (PLA mixed with graphene, G-PLA), and antenna support, made of transparent polymer (pure PLA). Selective galvanic metallization of antenna skeletons covers the later with mm-thick copper layers, sufficient to support high frequency electromagnetic currents. As a result, nontrivial antenna geometries have been demonstrated. Broadband and resonant directive antennas were characterized and exhibited performances, suitable for employing them in a wide range of GHz applications.

The developed methodology has many advantages over existing fabrication techniques. It allows exploiting complex volumetric geometries in antenna applications and provides guidelines for fast and unexpensive manufacturing. While direct metal printing techniques do exist, they still can demand usage of expensive machinery. Our methodology provides solutions, which can be adopted by end users without employing complex tools, and allows flexible custom-made designs.

## V. TECHNICAL INFORMATION

Archimedean spiral antenna parameters (in mm):

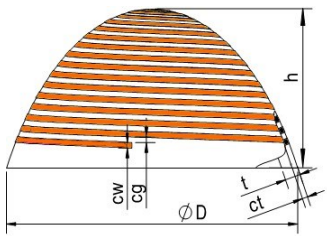

| D | Diameter | 103.87 |
|---|---|---|
| cg | conductive gap | 2 |
| ct | conductive thickness | 2 |
| cw | conductive width | 2 |
| h | height | 54 |
| t | thickness | 6 |

$$S_{prolate} = 2\pi a^2 \left(1 + \frac{c}{ae} \arcsin e \right) \quad \text{where} \quad e^2 = 1 - \frac{a^2}{c^2}$$

Yagi-Uda antenna parameters:

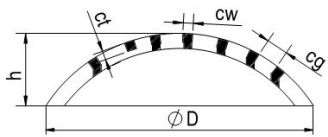

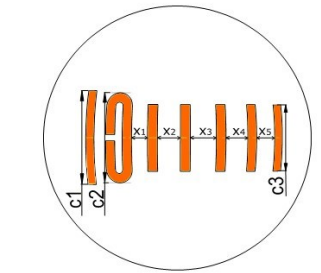

| cg | conductive gap | 7 |
|---|---|---|
| ct | conductive thickness | 3 |
| cw | conductive width | 3.5 |
| c1 | Reflector length | 32 |
| c2 | Folded dipole length | 30 |
| c3 | Director length | 22 |
| D | Diameter | 88 |
| h | Height | 24 |
| X1 | Folded dipole and director gap | 5.5 |
| X2 | First directors gap | 7.5 |
| X3 | Second directors gap | 8.5 |
| X4 | Third directors gap | 8 |
| X5 | Fourth directors gap | 7 |

Antenna connection layout:

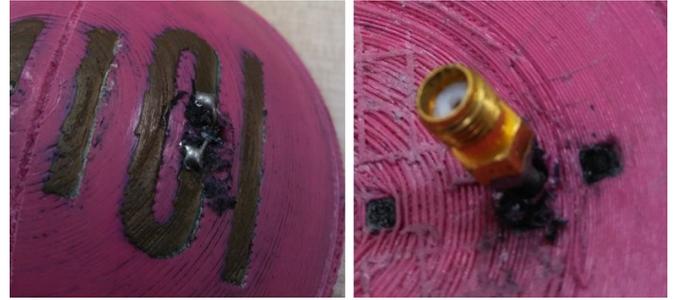

## VI. ACKNOWLEDGMENTS

The research was supported in part by PAZY Foundation and 3PEMS Ltd.